\documentstyle[12pt,epsfig]{aipproc} 
\newcommand{\vtd}     {\mbox{$V_{td}$}}
\newcommand{\vub}     {\mbox{$V_{ub}$}}
\newcommand{\vcb}     {\mbox{$V_{cb}$}}
\newcommand{\keiii}   {\mbox{$K^+ \! \rightarrow \! \pi^0 e^+ \nu_e$}}
\newcommand{\kmng}    {\mbox{$K^+ \! \rightarrow \! \mu^+ \nu_\mu \gamma$}}
\newcommand{\pme}     {\mbox{$\pi^+ \! \rightarrow \! \mu^+ \! \rightarrow \! e^+$}}
\newcommand{\kmn}     {\mbox{$K^+ \! \rightarrow \! \mu^+ \nu_\mu $}}
\newcommand{\kpp}     {\mbox{$K^+ \! \rightarrow \! \pi^+ \pi^\circ $}}
\newcommand{\kpnn}    {\mbox{$K^+ \! \rightarrow \! \pi^+ \nu \overline{\nu}$}}
\newcommand{\klpnn}   {\mbox{$K^\circ_L \! \rightarrow \! \pi^\circ \nu \overline{\nu}$}}
\title{Evidence for \kpnn}

\author{Steve Kettell\\
{\it Brookhaven National Laboratory}\\
{\it Upton, NY 11973} }

\date{}

\begin{document}

\begin{flushright}
{\underline{\bf BNL---66324 } \\
hep-ex/9903057\\
March 23, 1999}
\end{flushright}

\maketitle

\begin{abstract}
The first observation of the decay \kpnn\ has been reported.  The E787
experiment presented evidence for the \kpnn\ decay, based on the
observation of a single clean event from data collected during the
1995 run of the AGS (Alternating Gradient Synchrotron at Brookhaven
National Laboratory).  The branching ratio indicated by this
observation, B(\kpnn) = 4.2$^{+9.7}_{-3.5}\times10^{-10}$, is
consistent with the Standard Model expectation although the central
experimental value is four times larger.  The final E787 data sample,
from the 1995--98 runs, should reach a sensitivity of about five times
that of the 1995 run alone.  A new experiment, E949, has been given
scientific approval and should start data collection in 2001.  It is
expected to achieve a sensitivity of more than an order of magnitude
below the prediction of the Standard Model.
\end{abstract}

\section*{Introduction}

The rare decay \kpnn\ is a flavor changing neutral current, mediated
in the Standard Model (SM) by heavy quark loops\cite{GL,IL}, and is
sensitive to the Cabibbo-Kobayashi-Maskawa (CKM) matrix element $|\vtd
|$\cite{bf}.  This sensitivity arises from the heavy top quark mass,
the hard GIM suppression, and the negligible long distance
contribution\cite{hagelin,rein,lu,geng,faijfer}.  The hadronic matrix
element can be determined from the \keiii\ rate, with the inclusion of
isospin violating and electroweak radiative
effects\cite{marciano,bb97}.  The small QCD corrections have been
calculated to next to leading log\cite{bb94}. The QCD corrections to
the charm quark contribution are the major source of the intrinsic
theoretical uncertainty. This leads to a 7\% uncertainty in the
calculation of the branching ratio\cite{bf}.  If $B(\kpnn)$, $m_t$, $|
\vub / \vcb |$, and $|\vcb|$ were perfectly known, $|\vtd|$ could be
determined to $\sim$6\%. Given the current uncertainties in $m_t$,
$m_c$, \vcb, $|V_{ub}/V_{cb}|$, $\epsilon_K$ and $\bar B - B$ mixing,
the standard model prediction for the branching ratio, is
0.6---1.5$\times10^{-10}$.

A measurement of B(\kpnn) is one of the theoretically cleanest ways of
determining $| \vtd |$. Combined with the other `Golden Mode', \klpnn,
the CKM triangle can be completely determined from the $K$
system. With new measurements of the CKM parameters in the $B$ system
expected from the $B$-factories, additional tests of the Standard
Model, comparing results from $K$'s and $B$'s, will be possible. In
many extensions to the Standard Model the effects on the $K$ and $B$
system turn out to be discernibly different.

\section*{The E787 Experiment}

The experimental signature for \kpnn\ is a single incident kaon track
and a single outgoing pion track, with two missing neutrinos. The
separation of this signal from the background requires that the
particle identification and kinematics of the $\pi$ must be very well
measured and any additional particles must be vetoed with high
efficiency.

\begin{figure}[htb]
\center\epsfig{figure=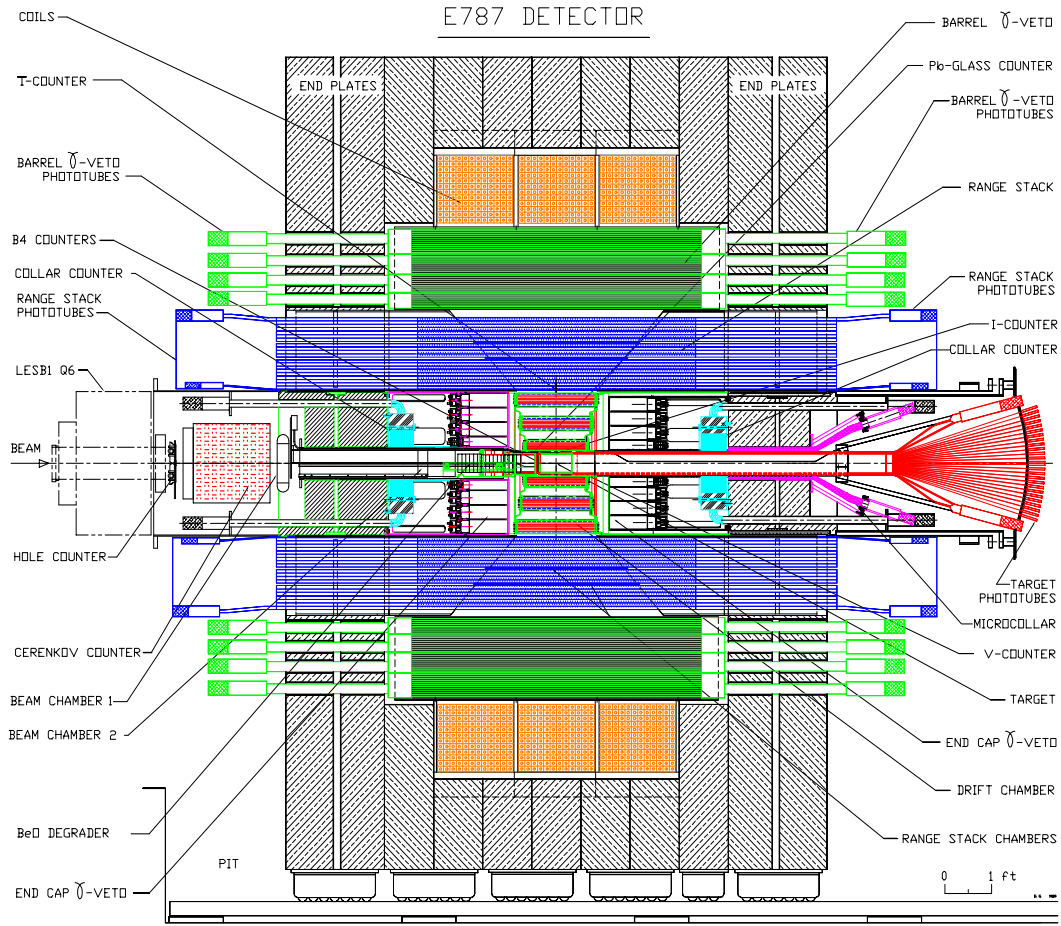,width=4.8in,height=5.5in,angle=90}
\caption{The E787 detector.}\label{f787_det}
\end{figure}

A drawing of the E787\cite{det} detector is shown in Fig.~\ref{f787_det}.
The detector is located in the C4 beam line at the AGS. This beam
line, Low Energy Separated Beam (LESB3)\cite{lesb3}, transports kaons
of up to 830 MeV/c. At 690 MeV/c it can transmit more than
$3\times10^6 K^+$/spill with a $K/\pi$ ratio of $>$3:1 for 10$^{13}$
protons on the production target.  The $K^+$ are tracked down the
beamline and stopped in a scintillating fiber target in the center of
the detector. The decay $\pi^+$ are tracked through the target and
drift chamber\cite{UTC} into the plastic scintillator range stack
(RS). The detector is located in a 1 T, solenoidal magnetic field.

The two most significant backgrounds are the two dominant $K^+$ decay
modes, \kpp\ (K$_{\pi2}$) and \kmn\ (K$_{\mu2}$), which produce
mono-energetic charged particles. The search region for \kpnn\
excludes these two kinematic peaks.  The E787 detector uses redundant
measures of the kinematics: momentum (P), energy (E) and range (R).
The K$_{\pi2}$ can also be suppressed by vetoing on the $\pi^0$
photons, so the detector is surrounded by a nearly 4$\pi$ photon veto
(PV). The primary components of the PV are the barrel veto (BV) and
endcap (EC)\cite{CsI1,CsI2,CsI3,CCD}, but, in fact, almost the entire
detector not traversed by the $\pi^{+}$ is used as a veto. The
K$_{\mu2}$ can also be suppressed by $dE/dx$ and by requiring that the
$\pi$ decay to a $\mu$ in the RS. The entire \pme\ decay chain is
observed with 500 MHz, 8-bit transient digitizers(TD)\cite{TD}
sampling the output of the RS scintillators.  The only other
significant background comes from scattered beam pions or $K^+$ charge
exchange (CEX).  The primary tools for rejecting these events are good
particle identification in the beam counters, identification of both
the $K^+$ and $\pi^+$ in the fiber target, and the requirement that
the $\pi^+$ track occur later than the $K^+$ track.

The search for \kpnn\ requires an identified $K^+$ to stop in the
target followed, after a delay of at least 2 ns, by a single
charged-particle track that is unaccompanied by any other decay
product or beam particle. This particle must be identified as a
$\pi^+$ with $P$, $R$ and $E$ between the $K_{\pi 2}$ and $K_{\mu 2}$
peaks.  To elude rejection, $K_{\mu 2}$ and $K_{\pi 2}$ events have to
be reconstructed incorrectly in $P$, $R$ and $E$. In addition, any
event with a muon has to have its track misidentified as a pion ---
the TD's provide a suppression factor $10^{-5}$.  Events with photons,
such as $K_{\pi 2}$ decays, are efficiently eliminated --- the
suppression of $\pi^0$'s is $10^{-6}$ (photon energy threshold of
$\sim$1 MeV). A scattered beam pion can survive the analysis only by
misidentification as a $K^+$ and if the track is mismeasured as
delayed, or if the track is missed entirely by the beam counters after
a valid $K^+$ stopped in the target. CEX background events can survive
only if the $K_L^0$ is produced at low enough energy to remain in the
target for at least 2 ns, if there is no visible gap between the beam
track and the observed $\pi^+$ track, and if the additional charged
lepton evades detection.

Reliable estimation of backgrounds is one of the most important
aspects of the measurement of \kpnn\ at the $10^{-10}$ level.  For
each source of background two independent sets of cuts are established
by taking advantage of the redundancy of detector measurements. One
set of cuts is relaxed or inverted to enhance the background (by up to
three orders of magnitude) so that the other group can be evaluated to
determine its power for rejection.  In this fashion backgrounds can be
studied at sensitivities up to 1000 times greater than the
experimental sensitivity. For example, $K_{\mu 2}$ (including \kmng)
are studied by separately measuring the rejections of the TD particle
identification and kinematic cuts.  The background from $K_{\pi 2}$ is
evaluated by separately measuring the rejections of the photon
detection and kinematic cuts.  The background from beam pion
scattering is evaluated by separately measuring the rejections of the
beam counter and timing cuts.  Extensive measurements of $K^+$ charge
exchange in the target were made in 1997.  This data combined with
Monte Carlo simulation of semi-leptonic $K_L$ decays, allows the CEX
background to be determined.  Small correlations in the separate
groups of cuts are investigated for each background source and
corrected for if they existed.

Further confidence in the background estimates and in the measurements
of the background distributions near the signal region is provided by
extending the method described above to estimate the number of events
expected with various degrees of cut loosening so as to allow higher
levels of all background types.  Confronting these estimates with
measurements from the full \kpnn\ data, where the two sets of cuts for
each background type were relaxed simultaneously, tested the
independence of the two sets of cuts.  The background level for the
1995 data set, $b$, was measured to be 0.08 events. At the level of
$20 \times b$, two events were observed where $1.6\pm0.6$ were
expected, and at $150 \times b$, 15 events were found where $12 \pm 5$
were expected.  Under detailed examination, the events admitted by the
relaxed cuts were consistent with being due to the known background
sources.  Within the final signal region, additional background
rejection capability is available.  Therefore, prior to looking in the
signal region, several sets of increasingly tighter criteria were
established, to be used only to interpret any events in the signal
region.

\section*{The Event}

\begin{figure}[htb]
\begin{minipage}[htb]{.44\linewidth}
\center\epsfig{figure=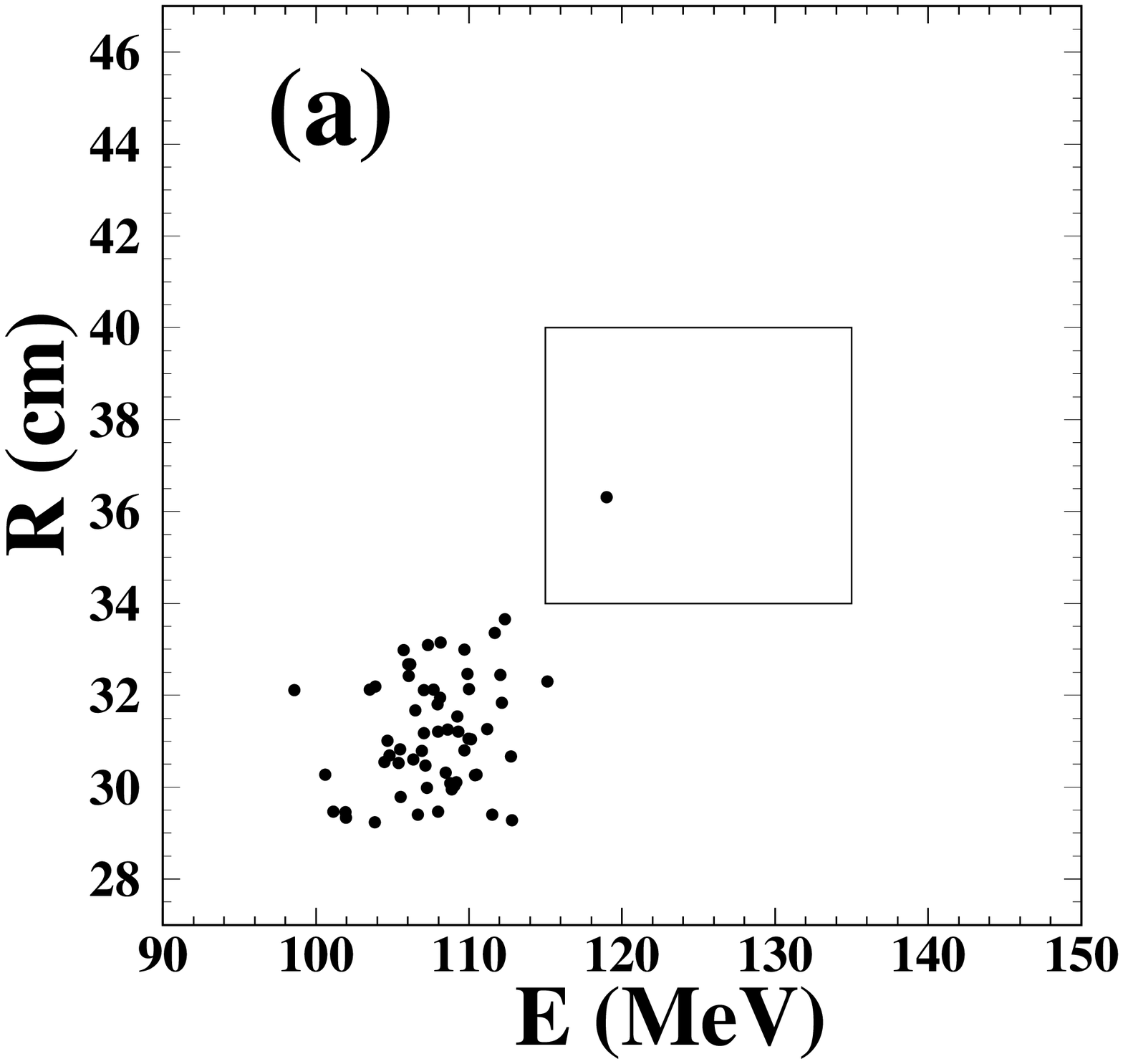,width=\linewidth,height=2.5in,angle=0}
\end{minipage}\hfil
\begin{minipage}[htb]{.44\linewidth}
\center{\epsfig{figure=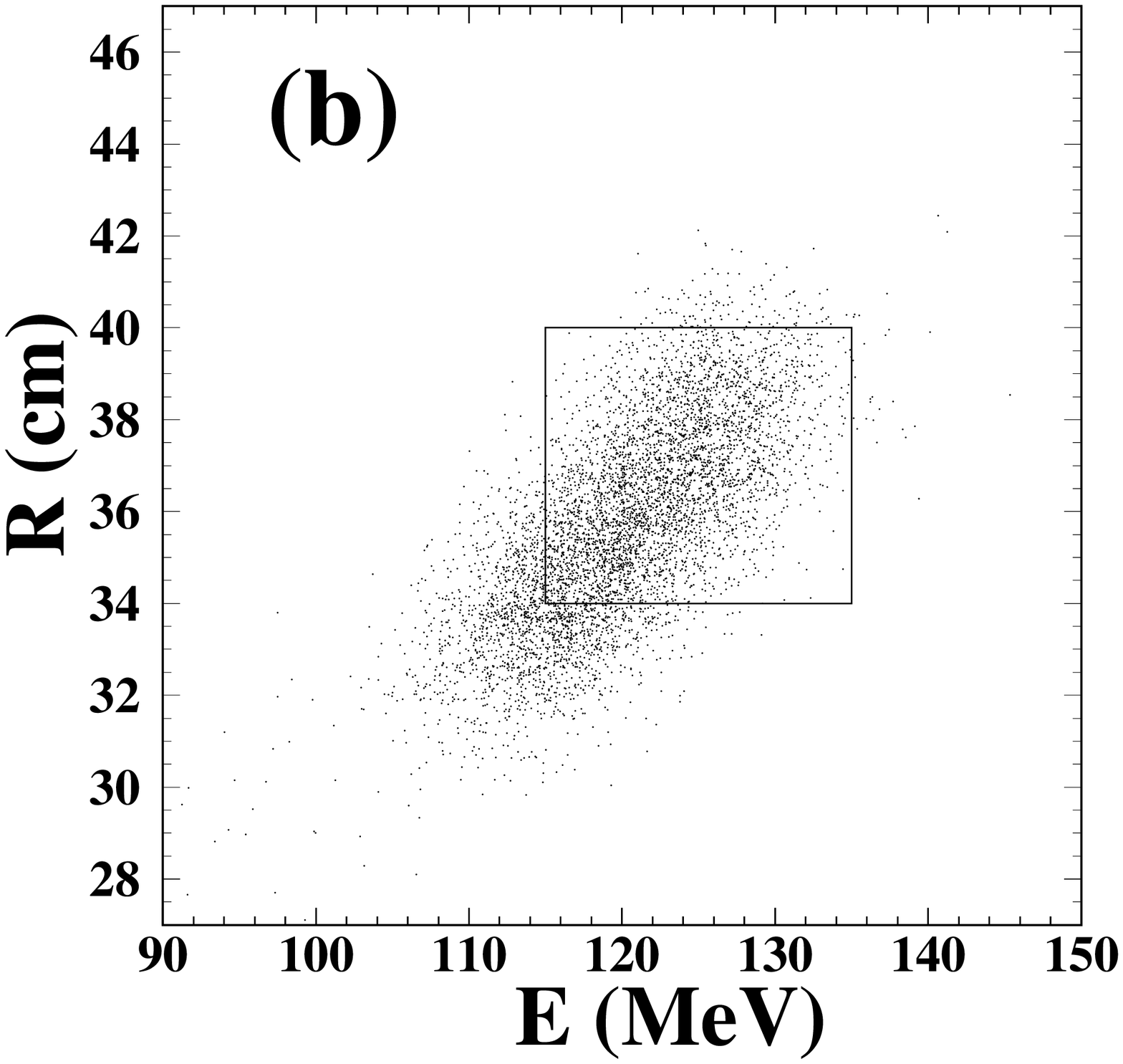,width=\linewidth,height=2.5in,angle=0}}
\end{minipage}
{\caption{\label{787_pnn1}Final event candidate for \kpnn: a) Data b)
Monte-Carlo} }
\end{figure}

\begin{figure}[htb]
\center\epsfig{figure=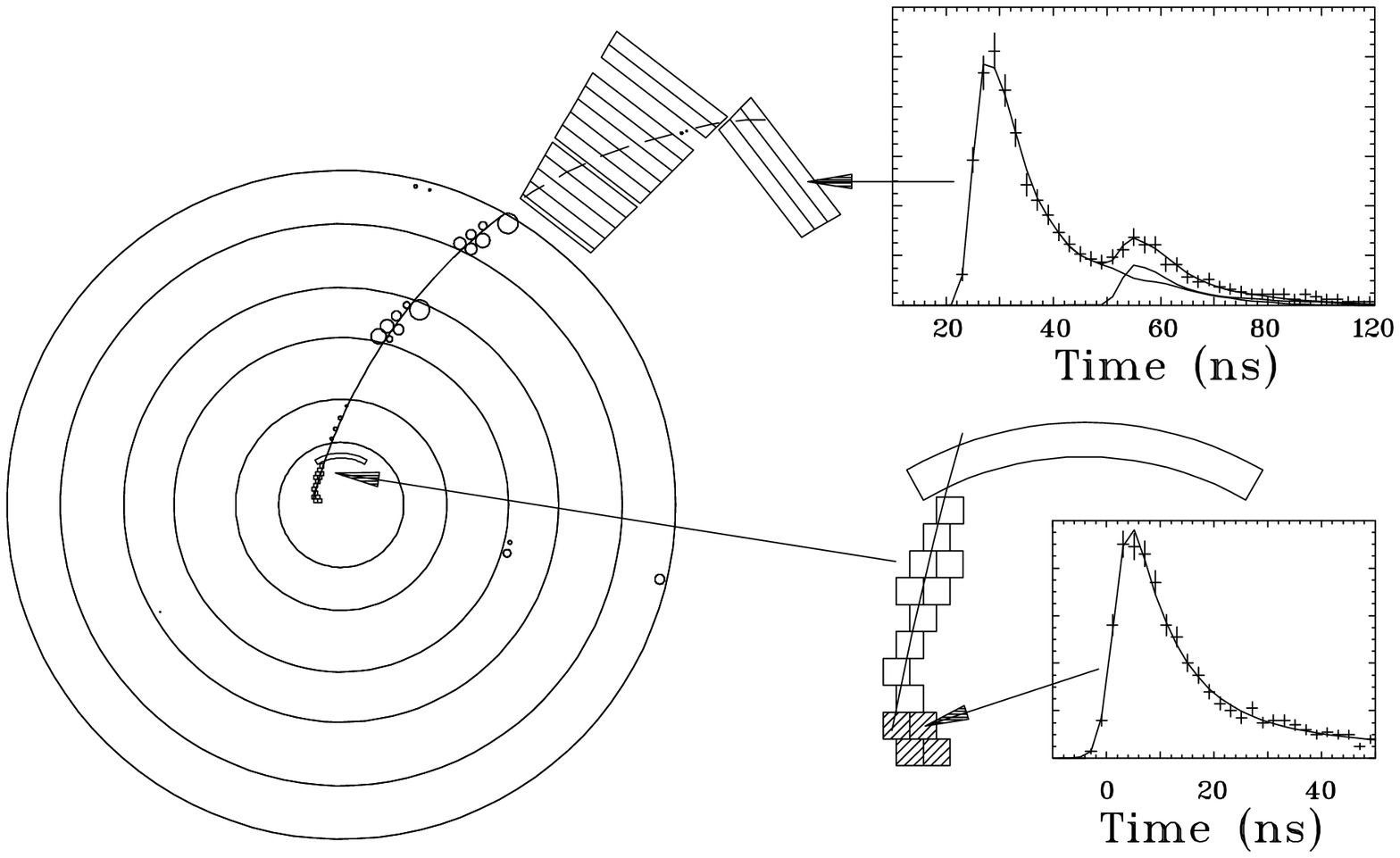,width=4.5in,height=2.75in,angle=0}
\caption{The first \kpnn\ event.}\label{787_pnn2}
\end{figure}

E787 published the first observation of the \kpnn\ decay\cite{pnn6},
based on data from the 1995 run of the AGS.  The range and energy of
event candidates passing all other cuts is shown in
Fig.~\ref{787_pnn1}. The box (which encloses the upper 16.2\% of the
\kpnn\ phase space) indicates the signal region.
One event consistent with the decay \kpnn\ was observed. The expected
number of background events from all sources was $0.08\pm0.03$ events
(branching ratio equivalent of $3\times10^{-11}$). The separate levels
of background were: K$_{\mu2}$=$0.02\pm0.02$,
K$_{\pi2}$=$0.03\pm0.02$, beam-$\pi^+$=$0.02\pm0.01$ and
CEX=$0.01\pm0.01$.  This event was in a particularly clean region
where the expected background was $0.008\pm0.005$ and which contained
55\% of the acceptance of the full signal region. A reconstruction of
the event is shown in Fig.~\ref{787_pnn2}. The kaon decayed to a pion
at 23.9 ns, followed by a clean $\pi^+\rightarrow\mu^+$ decay 27.0 ns
later, as can be seen in the upper insert in Fig.~\ref{787_pnn2};
there was also a clean $\mu^+\rightarrow e^+$ decay at 3201.1 ns.

There was no significant activity anywhere else in the detector at the
time of the $K^+$ decay. The lower insert in Fig.~\ref{787_pnn2} shows
one of the target fibers that contains the incident kaon track, yet
does not lie on the outgoing $\pi^+$ track; there is no activity at
the $K^+$ decay time.
The branching ratio for \kpnn\ implied by the observation of this
event is B(\kpnn) = 4.2$^{+9.7}_{-3.5} \times 10^{-10}$.

\section*{Expectations for the E787 Final Result}

The E787 experiment has had four runs during 1995--98.  The typical
conditions for the 1995 run were 13 Tp/spill, 5.3 MHz of incident
$K^+$, a stopped kaon rate of 1.2 M/spill, a deadtime of 25\%, and an
acceptance of 0.16\%.  The acceptance has been measured to be 60\% of
the acceptance at zero rate.  The rates in most detector elements have
been measured to be proportional to the incident flux and not to the
stopped kaon flux.  This implies that the sensitivity increases with
increasing fraction of stopped kaons/incident kaons. Therefore E787
has lowered the momentum of the incident kaons in subsequent years,
increasing the sensitivity without increasing the rates in most
detector elements. In addition, the duty factor of the AGS has been
increased from 37\% at the start of 1995 to 55\% at the end of 1998.
Improvements in the online trigger efficiency, live time from trigger
and DAQ upgrades\cite{e787_daq1,e787_daq2,e787_daq3,e787_daq4}, and
acceptance from running at lower momentum also contribute to the
enhanced sensitivity in later years.

A summary of the E787 sensitivity, including the published 1995
result, is shown in Table~\ref{tab_sens}, along with some of the
running conditions.
\begin{table}
\caption{Running conditions for E787. The number of stopped kaons with
the detector live(KB$_{L}$) is the online measure of the experimental
sensitivity. As the kaon momentum is lowered the fraction of incident
kaons that stop in the target(sf) increases. The duty factor(DF) of
the AGS steadily increased. The single event sensitivity(S.E.S.) is
measured from the 1995 set and estimated for the later years based on
KB$_{L}$ and the acceptance calculated from the detector rates. The
background levels (bck) for 1996--98 are estimated from the reanalysis
of the 1995 data.}\label{tab_sens}
\begin{center}
\begin{tabular}{|l||l|c|r|c||l||l|} \hline
{\bf year}&{\bf\boldmath KB$_{L}$}&{\bf\boldmath $|\vec{p_K}|$}&{\bf DF}&{\bf sf}&{\bf (S.E.S.)$^{-1}$}&{\bf bck}\\ 
&{\bf ($10^{12}$)}& &{\bf(\%)}&{\bf (\%)}&{\bf (10$^{10}$)}&{\bf events}\\ \hline\hline
       1995 & 1.49 & 790 & 41 & 18.7 & 0.24 & 0.08 \\ \hline\hline
       1995--97 & 3.33 & 670--790 & 43 & 22 & 0.5 & 0.09 \\ \hline
       1998     & 2.97 & 710 & 52 & 27 & 0.5 & 0.08 \\ \hline\hline
       1995--98 & 6.3 & 670--790 & 47 & 24 & 1.1 & 0.16 \\ \hline
\end{tabular}
\end{center}
\end{table}

 The expected sensitivity from the 1995--97 runs is $\sim\times$2.3
that of the 1995 data alone, even though these runs were considerably
shorter.  A preliminary re-analysis of the E787 1995 data with
improvements in the analysis software have demonstrated a background
rejection that is $\sim\times$2.3 larger.  This background level
(roughly equivalent to a branching ratio of $1.5\times10^{-11}$) is
sufficient for future measurements of the \kpnn\ branching ratio.
Results of the analysis of the larger data set are expected within a
few months.

With the improved running conditions, including an increased duty
factor, improved DAQ\cite{e787_daq5} and the relatively long running
period of almost five months in 1998, the final E787 sensitivity for
\kpnn\ should extend below the most probable SM level
($\sim1\times10^{-10}$).

\section*{Future Plans --- E949}

A new experiment to measure the branching ratio B(\kpnn),
E949\cite{949}, recently received scientific approval and is expected
to run at the AGS starting in the year 2001. This experiment is
designed to reach a sensitivity of (8--14)$\times10^{-12}$, an order
of magnitude below the Standard Model prediction and to determine
$|\vtd|$ to better than 27\%. It is built around the existing E787
detector to take advantage of the extensive analysis of that detector,
allowing a reliable projection of the new experiment to the required
sensitivity with a high level of confidence.

The E949 detector will have significantly upgraded photon veto
systems, DAQ and trigger compared to the E787 experiment.  The PV
upgrade includes a barrel veto liner that will replace the outer
layers of the RS and fill a gap between the RS and BV.  It is 2.3
$X_\circ$ thick and will add substantially to the thin region at
45$^\circ$. Additional PV upgrades will be installed along the beam
direction. The most important DAQ upgrade will be to instrument the RS
with TDC's to extend the search time for the Michel electron
($\mu^+\rightarrow e^+$) and to allow the TD range to be
shortened. The shortening of the TD range should allow a reduction of
deadtime by 30--50\%. Trigger upgrades should reduce the deadtime
further and reduce the acceptance loss due to the online PV, by
improving the timing on the RS and BV.  Compared to the E787 running
conditions in 1995 an improvement of 50\% has already been
realized. Additional improvements in these areas and in offline
software are expected to gain another 90\%. Additional sensitivity
gains can be realized by including the region of phase space below the
K$_{\pi2}$ peak and by reoptimizing the analysis algorithms to run at
higher rates.  Each of these should provide a factor of 2 more
sensitivity. The proposal assumes that only one of these factors will
be realized.

The operating conditions will be significantly upgraded.  E949 will
run with a 700 MeV/c $K^+$ beam, with more than 70 Tp
($7\times10^{13}$ protons per spill) and with an AGS duty factor of
close to 70\%.  These conditions are all within the expected AGS
operating parameters for the year 2000\cite{ags_beam}.  They will
require 
magnets in the LESB3 beam line, which are already near the end of
their lifetime from radiation damage.  The gain in sensitivity from
these conditions will be a factor of 2.2.  A summary of the
improvement factors is given in Table~\ref{tab_sens_factors}.
\begin{table}
\caption{Sensitivity improvement factors for E949,
compared to the published E787 result. }\label{tab_sens_factors}
\begin{center}
\begin{tabular}{|l||c|} \hline
{\bf Upgrade}           & {\bf Improvement factor} \\ \hline\hline
Lower momentum          &       1.44    \\ \hline
Higher duty factor      &       1.53    \\ \hline
Established improvements       &       1.54    \\ \hline
Additional efficiency improvements      &       1.9     \\ \hline
Phase space below \kpp  &       2       \\ \hline\hline
Total                   &       13      \\ \hline
\end{tabular}
\end{center}
\end{table}
The total gain in sensitivity per hour will be 6--13 times over the
E787 published result on the 1995 data set. All of the measurements of
deadtime and acceptance as a function of rate; and of the stopping
fraction and kaon flux have been included in a calculation of the
optimum running conditions. The sensitivity of the 1995 conditions is
$1.46\times10^6$/hr and for E949 is $9.6\times10^6$/hr, not including
the plans for reoptimizing for higher rates or the phase space below
the K$_{\pi2}$ peak.

\section*{Conclusions}

\begin{figure}[htb]
\center\epsfig{figure=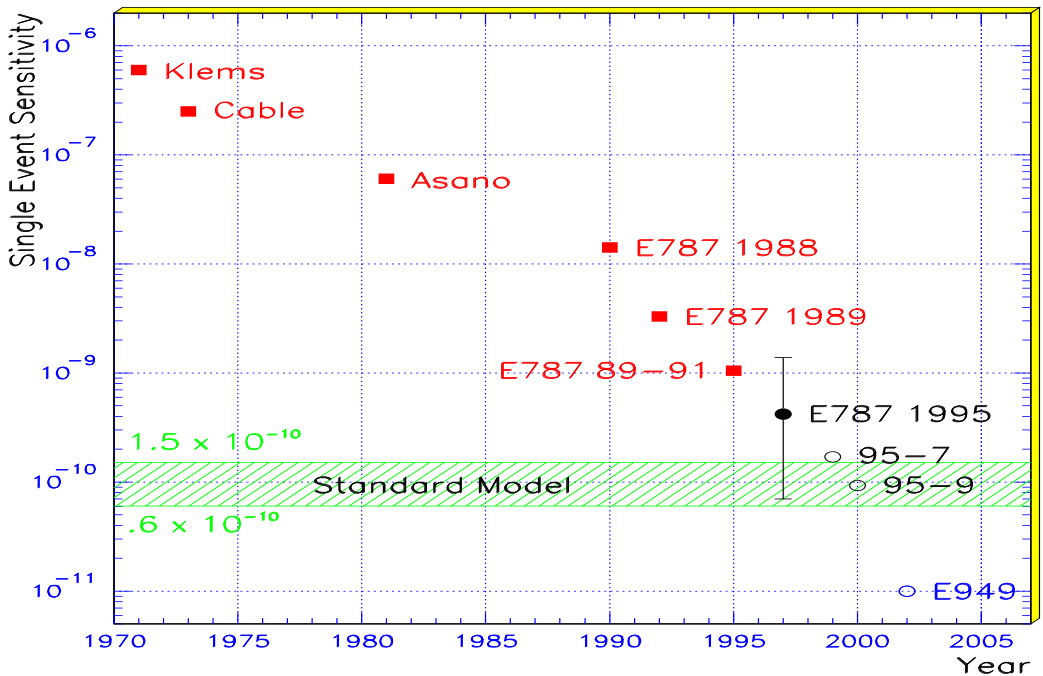,width=6.0in,height=4.in,angle=0}
\caption{History of progress in the search for \kpnn.  The sensitivity
of experiments setting limits is shown in solid squares. The first
actual measurement is shown as a solid circle and the projected future
measurements are shown as open circles. The background levels are
shown as stars for the recent data.}\label{fig_brt}
\end{figure}

The prospects for further improvement in the determination of B(\kpnn)
are bright. The first observation of this rare and interesting decay
has recently been published. The data on hand, or soon to be
available, from the E787 experiment, should provide almost an order of
magnitude more sensitivity. The recently approved experiment E949
should reach at least a factor of five further than E787 and make a
very interesting measurement of $| \vtd |$.  There is also a proposal,
CKM, at the FNAL Main Injector, to push even further, to $10^{-12}$ by
looking for the decay in flight.  A plot showing the progress from
past, current and approved experiments for B(\kpnn) is shown in
Figure~\ref{fig_brt}.
The search for this decay, with its very clean and well understood
theory, has had a long and now very fruitful history.

\section*{Acknowledgements}

Members of the E787/E949 collaborations:

\begin{itemize}
  \item {\bf University of Alberta}: P.~Kitching, H.-S.~Ng, R.~Soluk
  \item {\bf Brookhaven National Laboratory}: S.~Adler, M.S.~Atiya,
  I-H.~Chiang, M.V.~Diwan, J.S.~Frank, J.S.~Haggerty, V.~Jain,
  S.H.~Kettell, T.F.~Kycia,  K.K.~Li, L.S.~Littenberg, C.~Ng,
  A.~Sambamurti, A.J.~Stevens, R.C.~Strand, C.~Witzig 
  \item {\bf Fukui University}: M.~Miyajima, J.~Nishide, K.~Shimada, 
  T.~Shimoyama, Y.~Tamagawa 
  \item {\bf INR}: M.P.~Grigoriev, A.P.~Ivashkin, M.M.~Khabibullin, 
  A.N.~Khotjantsev, Y.G.~Kudenko, O.V.~Mineev
  \item {\bf KEK---Tanashi}: T.K.~Komatsubara, M.~Kuriki,
  N.~Muramatsu, K.~Omata, S.~Sugimoto 
  \item {\bf KEK---Tsukuba}: M.~Aoki, T.~Inagaki, S.~Kabe, M.~Kobayashi, 
  Y.~Kuno, T.~Sato, T.~Shinkawa, Y.~Yoshimura 
  \item {\bf Osaka University}: Y.~Kishi, T.~Nakano, M.~Nomachi 
  \item {\bf Princeton University}: M.~Ardebili,
  A.O.~Bazarko, M.R.~Convery, M.M.~Ito, D.R.~Marlow,
  R.A.~M\raisebox{.5ex}{c}Pherson, P.D.~Meyers, F.C.~Shoemaker,
  A.J.S.~Smith, J.R.~Stone 
  \item {\bf TRIUMF}: P.C.~Bergbusch,
  E.W.~Blackmore, D.A.~Bryman, S.~Chen, A.~Konaka, J.A.~Macdonald,
  J.~Mildenberger, T.~Numao, P.~Padley, J.-M.~Poutissou, R.~Poutissou,
  G.~Redlinger, J.~Roy, A.S.~Turcot
\end{itemize}

This work was supported under U.S. Department of Energy contract 
\#DE-AC02-98CH10886.


\begin{thebibliography}{99}
\bibitem{GL}M.K.~Gaillard~and~B.W.~Lee, {\it Phys. Rev.} {\bf D10}, 897 (1974).
\bibitem{IL}T. Inami and C. S. Lim, {\it Prog. Theor. Phys.} {\bf 65}, 297 
(1981).
\bibitem{bf}A.J.~Buras and R.~Fleischer, {\it Heavy Flavours II}, World 
Scientific, 1998, ed. Buras and Linder, p65; hep-ph/9704376.
\bibitem{hagelin}J.S.~Hagelin and L.S.~Littenberg, {\it Prog. Part. Nucl. 
Phys.}  {\bf 23}, 1 (1989).
\bibitem{rein}D.~Rein and L.M.~Sehgal, {\it Phys. Rev.} {\bf D39}, 3325 (1989).
\bibitem{lu}M.~Lu and M.B.~Wise, {\it Phys. Lett.} {\bf B324}, 461 (1994); 
hep-ph/9401204.
\bibitem{geng}C.Q.~Geng {\it et al.}, {\it Phys. Lett.} {\bf B355}, 569 (1995);
 hep-ph/9506313.
\bibitem{faijfer}S.~Faijfer, {\it Nuovo. Cim.} {\bf 110A}, 397 (1997).
\bibitem{marciano}W.J.~Marciano~and~Z.~Parsa, {\it Phys. Rev.} {\bf D53}, R1 
(1996).
\bibitem{bb97}G.~Buchalla~and~A.J.~Buras, {\it Phys. Rev.} {\bf D57},
216 (1998); hep-ph/9707243.
\bibitem{bb94}G. Buchalla and A.J. Buras, {\it Nucl. Phys.} {\bf B412}, 106 
(1994); hep-ph/9308272.
\bibitem{det}M.S.~Atiya {\it et al.}, NIM {\bf A321}, 129 (1992).
\bibitem{lesb3}I.-H.~Chiang {\it et al.}, to be submitted to NIM {\bf A}.
\bibitem{UTC}E.W.~Blackmore {\it et al.}, {\it NIM} {\bf A404}, 295 (1998).
\bibitem{CsI1}I.-H.~Chiang {\it et al.}, {\it IEEE Trans. Nucl. Sci.} 
{\bf NS--42}, 394 (1995).
\bibitem{CsI2}T.K.~Komatsubara {\it et al.}, {\it NIM} {\bf A404}, 315 (1998).
\bibitem{CsI3}M.~Kobayashi {\it et al.}, {\it NIM} {\bf A337}, 355 (1994).
\bibitem{CCD}D.A.~Bryman {\it et al.}, {\it NIM} {\bf A396}, 394 (1997).
\bibitem{TD}M.S.~Atiya {\it et al.}, {\it NIM} {\bf A279}, 180 (1989).
\bibitem{pnn6}S.~Adler {\it et al.}, {\it Phys. Rev. Lett.} {\bf 79},
2204 (1997); hep-ex/9708031.
\bibitem{e787_daq1}M.~Burke  {\it et al.}, {\it IEEE Trans. Nucl. Sci.} 
{\bf NS--41}, 131 (1994).
\bibitem{e787_daq2}C.~Witzig and S.~Adler, {\it Real-Time Comput. Appl.}, 
123 (1993).
\bibitem{e787_daq3}S.S.~Adler, {\it Inter. Conf. Electr. Part. Phys.}, 133 
(1997).
\bibitem{e787_daq4}C.~Zein {\it et al.}, {\it Real-Time Comput. Appl.}, 103 
(1993).
\bibitem{e787_daq5}S.S.~Adler {\it et al.}, to be submitted to IEEE.
\bibitem{949}M.~Aoki {\it et al.}, {\it AGS Proposal 949}, August 1998.
\bibitem{ags_beam}J.M. Brennan and T. Roser, {\it 'High intensity performance 
of the Brookhaven AGS'}, 5$^{th}$ Europ.Part.Acc.Conf.(EPAC96), 530 (1996).
\end{thebibliography}
\end{document}